\documentclass[]{an}
\usepackage{graphicx,times,fancyhdr}
\usepackage{psfig}
\usepackage{afterpage}
\begin{document}


\renewcommand{\vec}[1]{\mbox{\boldmath $#1$}}
\def\bl{\par\vskip 12pt\noindent}

\def\Pm{\mathop{\rm Pm}\nolimits}
\def\qq{\qquad\qquad}                      
\def\q{\qquad}
\def\bib{\item}
\def\beg{\begin{eqnarray}}
\def\ende{\end{eqnarray}}
\def\Om{{\it \Omega}}
\def\Del{{\it \Delta}}
\def\gsim{\lower.4ex\hbox{$\;\buildrel >\over{\scriptstyle\sim}\;$}}
\def\lsim{\lower.4ex\hbox{$\;\buildrel <\over{\scriptstyle\sim}\;$}}

\title{The stability of MHD Taylor-Couette flow with  current-free spiral
 magnetic fields between  conducting cylinders}
\author{G. R\"udiger\inst{1,2} \and
R. Hollerbach\inst{1,3} \and
M. Schultz\inst{2} \and
D.A. Shalybkov\inst{2,4}}
\institute{Isaac Newton Institute for Mathematical Sciences, 20 Clarkson Rd., Cambridge CB3 0EH, UK \and Astrophysikalisches Institut Potsdam,
         An der Sternwarte 16, 14482 Potsdam, Germany \and
Department of Mathematics, University of Glasgow, Glasgow G12 8QW,
UK
\and     
A.F. Ioffe Institute for Physics and Technology,
          St. Petersburg 194021, Russia}

\abstract{
We study the magnetorotational instability in cylindrical Taylor-Couette flow,
with the (vertically unbounded) cylinders taken to be perfect conductors, and
with externally imposed spiral magnetic fields.  The azimuthal component of this
field is generated by an axial current inside the inner cylinder, and may be
slightly stronger than the axial field.  We obtain an instability beyond
the Rayleigh line, for Reynolds numbers of order 1000 and Hartmann numbers of
order 10, and independent of the (small) magnetic Prandtl number.
For  experiments with $R_{\rm out}=2\ R_{\rm in} =10$ cm and $
\Omega_{\rm out} = 0.27\ \Omega_{\rm in}$, the instability appears for liquid
sodium for axial fields of $\sim$20 Gauss and axial currents of $\sim$1200 A.
For gallium the numbers are $\sim$50 Gauss and $\sim$3200 A.
The vertical cell size
is about twice the cell size known for nonmagnetic experiments.
}

\correspondence{gruediger@aip.de
}
\maketitle

\section{Motivation}
A limited number of instabilities are responsible for most of the pattern
formation in the Universe.  Stars are formed by the Jeans instability, and
the heat transport within them is driven by the
Rayleigh-B\'{e}nard instability.  Most high-energy radiation, however, is
produced by disks around compact objects and black holes.  These accretion
disks must therefore be turbulent.  It is now believed that this turbulence
is caused by the instability of their Keplerian rotation law in the presence
of weak large-scale magnetic fields.  This `magnetorotational instability' (MRI)
was first discovered by Velikhov (1959), who considered an electrically
conducting fluid between rotating cylinders with $\Om_{\rm out} < \Om_{\rm in}$.
The existence of this instability has been confirmed analytically
and numerically many times; see for example Balbus (2003) and R\"udiger \& Hollerbach (2004)  for  recent reviews.

There is considerable interest in studying the magnetorotational instability
in the laboratory. The simplest, and most widely studied design, is to confine
a liquid metal between differentially rotating cylinders, and impose a magnetic
field along the axis of the cylinders.  This yields the MRI for magnetic
Reynolds numbers of order 10.  However, because of the extremely small
magnetic Prandtl numbers $\rm Pm(=\nu/\eta)$  of available liquid metals (see Table \ref{tab1}), that translates into
ordinary Reynolds numbers exceeding $10^6$, which causes severe difficulties
(Hollerbach \& Fournier 2004).  In order to overcome such problems,
Hollerbach \& R\"udiger (2005) proposed
imposing an azimuthal field as well, with dramatic consequences, namely a
reduction of the critical Reynolds number $\rm Re_c$ from $O(10^6)$ to $O(10^3)$.  These new solutions are
also essentially independent of Pm, and are therefore ideally suited to
experimental realisations in the laboratory.  These results were for
insulating inner and outer cylinders, in which case the axial current one
must impose to generate this azimuthal magnetic field is around 2500 A 
(using liquid sodium as the fluid), which is close to the upper limit of what
is experimentally possible.  Here we therefore consider conducting cylinders,
and show that the necessary current (as well as the Reynolds number)
is smaller than for insulating cylinders.
\begin{table}[htb]
\tabcolsep5pt
\caption{\label{tab1} Liquid metal material parameters in c.g.s.
}
\begin{tabular}{l|ccccc}
\hline
 & $\rho$  & $\nu$  & $\eta$ & Pm & $\sqrt{\mu_0 \rho
 \nu \eta}$\\[0.5ex]
\hline\\[-8pt]
sodium & 0.9 & $7.1\cdot 10^{-3}$ & $0.8\cdot 10^3$ &$0.9\cdot 10^{-5}$ &8.15\\[0.5ex]
 gallium & 6.0 & $3.2\cdot 10^{-3}$ & $2.1\cdot 10^3$ &$1.5 \cdot 10^{-6}$ &22.0\\[0.5ex]
    \hline
 \end{tabular}
 \end{table}
\section{Introduction}
\begin{figure}[htb]
\hskip1.5cm 
\psfig{figure=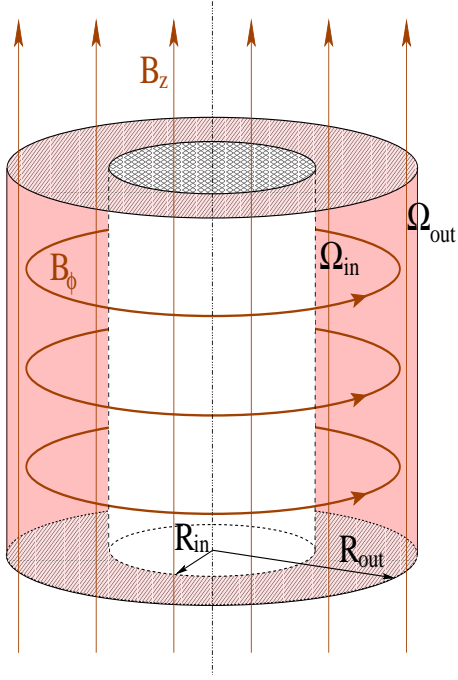,width=6.0cm,height=9.0cm}  
\caption{(online colour at www.an-journal.org)
The basic geometry of the problem, consisting of two cylinders of radii
$R_{\rm in}$ and $R_{\rm out}$, rotating at $\Om_{\rm in}$ and $\Om_{\rm out}$.
$B_z$ and $B_\phi$ are the externally imposed magnetic fields, $B_\phi$ by an
axial current inside the inner cylinder.}
\label{geo}
\end{figure}

We consider the stability of the flow between two rotating coaxial infinitely
long cylinders, in the presence of a constant axial field $B_z$, and a 
current-free (within the fluid) azimuthal field $B_\phi$.  Figure~\ref{geo}
shows a sketch of the geometry.  The fluid is incompressible, with density
$\rho$, kinematic viscosity $\nu$ and magnetic diffusivity $\eta$.

By conservation of angular momentum the basic state differential rotation
profile is given by
\begin{equation}
\Om=a+\frac{b}{R^2},
\label{1}
\end{equation}
where $a$ and $b$ are given by
\begin{equation}
a= \frac{\hat\mu-\hat{\eta}^2}{1-\hat{\eta}^2} \Om_{\rm in},
\q b= \frac{1-\hat\mu}{1-\hat{\eta}^2} R_{\rm in}^2 \Om_{\rm in},
\label{2}
\end{equation}
with $\hat\mu=\Om_{\rm out}/\Om_{\rm in}$ and
$\hat\eta=R_{\rm in}/R_{\rm out}$. $R_{\rm in}$ and $R_{\rm out}$ are the
radii of the inner and outer cylinders, and $\Om_{\rm in}$ and $\Om_{\rm out}$
are their angular velocities.

>From the azimuthal component of the induction equation one finds
\begin{equation}
B_\phi=\frac{B}{R}
\label{3}
\end{equation}
with 
$
B=\beta B_0 R_{\rm in}.
$
This parameter $\beta$ therefore denotes the ratio of the toroidal 
field to the constant axial field $B_z=B_0$. This toroidal field $B_\phi$ is
maintained by an electric current running along the central axis of strength
\begin{equation}
J=5\beta B_0 R_{\rm in}
\label{J}
\end{equation}
with $J$ in Ampere, $B_0$ in Gauss and $R_{\rm in}$ in cm.

We are interested in the stability of the basic state (\ref{1})
against axisymmetric and nonaxisymmetric perturbations.
The perturbed state of the flow is  described by the quantities
\begin{equation}
u'_R, \ \ R\Om+u'_\phi, \ \ u'_z,   \ \ B'_R, \ \ B_\phi+B'_\phi, \ \ B_0+B'_z.
\end{equation}
The solutions
of the linearized MHD equations are considered in their modal representation
$\vec{F}'=\vec{F}'(R){\textrm{exp}}({\textrm{i}}(kz+m\phi+\omega t))$ where $\vec{F}'$ is any
of $\vec{u}'$ and $\vec{B}'$.
The dimensionless numbers of the problem are the magnetic Prandtl number Pm,
the Hartmann number Ha and the Reynolds number Re,
\begin{equation}
{\textrm{Pm}} = \frac{\nu}{\eta},\quad
{\textrm{Ha}}=\frac{B_{0} R_0}{\sqrt{\mu_0 \rho \nu \eta}},\quad 
{\textrm{Re}}=\frac{\Om_{\textrm{in}} R_0^2}{\nu},
\label{pm}
\end{equation}
where $\mu_0$ is the permeability, and 
$R_0=(R_{\rm in}D)^{1/2}$ with $D=R_{\rm out}-R_{\rm in}$. 

In order to understand the results presented here, it is useful also to
consider the induction equation for the fluctuating toroidal field component,
\begin{equation}
 \frac{\partial B_\phi'}{\partial t}-\eta \Del B_\phi'=  
  {\rm
rot}_\phi\big(\bar{\vec{u}}_\phi \times \vec{B}_{\rm pol}'+ \vec{u}'_\phi\times
\bar{\vec{B}}_{\rm pol} + \vec{u}_{\rm pol}'\times \bar{\vec{B}}_\phi\big)
\label{torfield}
\end{equation}
where $\vec u_{\rm pol}$ and $\vec B_{\rm pol}$ denote the poloidal components of $\vec u$ and $\vec B$. Chandrasekhar (1961) cancelled the last two terms on the right of this 
equation, and did not obtain the MRI. It occurs if only the last term
in (\ref{torfield}) is cancelled; but one can show that for small Pm it only
appears for ${\rm Rm}=O(10)$. If also the last term in (\ref{torfield}) is
retained then the resulting Reynolds number loses its strong  ${\rm
Re}\propto{\rm Pm}^{-1}$ dependence, and is reduced by 3 orders of
magnitude (Hollerbach \& R\"udiger 2005, see also Table \ref{tabold}).

\section{The equations}
The equations of  the problem are given in a similar form as by
R\"udiger \& Shalybkov (2004). Lengths and wave numbers are normalized by
$R_0$, velocities by $\eta/R_0$, frequencies by $\Om_{\rm in}$, and magnetic
fields by $B_0$. One then obtains
\begin{eqnarray}
&&\frac{{\rm d}P'}{{\rm d}R}+{\rm i}\frac{m}{R}X_2+{\rm i}kX_3+\left(k^2+\frac{m^2}{R^2}
\right)u'_R+\nonumber\\
&& \q +{\rm iRe}(\omega+m\Om)u'_R-2\Om\,{\rm Re}\,u'_\phi -
{\rm iHa}^2kB'_R-\nonumber\\
&& \q -{\rm iHa}^2\frac{mB}{R^2} B'_R +2{\rm Ha}^2\frac{B}{R^2}
B'_\phi=0,
\label{dpdr}
\end{eqnarray}
\begin{eqnarray}
&& \frac{{\rm d}X_2}{{\rm d}R}-\left(\!k^2+\frac{m^2}{R^2}\!\right)u'_\phi-
{\rm iRe}(\omega+m\Om)u'_\phi+2{\rm i}\frac{m}{R^2}u'_R\!-\nonumber\\
&& -2{\rm Re}\,a\,u'_R
+{\rm iHa}^2\frac{mB}{R^2} B'_\phi + {\rm iHa}^2kB'_\phi
-{\rm i}\frac{m}{R}P'=0,
\label{dx2dr}
\end{eqnarray}
\begin{eqnarray}
&& \frac{{\rm d}X_3}{{\rm d}R}+\frac{X_3}{R}-\left(k^2+\frac{m^2}{R^2}\right)u'_z-
{\rm iRe}(\omega+m\Om)u'_z-\nonumber\\
&& \qq -{\rm i}kP'+{\rm iHa}^2\frac{mB}{R^2} B'_z
+{\rm iHa}^2kB'_z=0,
\label{dx3dr}
\end{eqnarray}

\beg
&& \frac{{\rm d}u'_R}{{\rm d}R}+\frac{u'_R}{R}+{\rm i}\frac{m}{R}u'_\phi+{\rm i}ku'_z=0,
\label{5}
\ende
\begin{eqnarray}
&&\frac{{\rm d}B'_z}{{\rm d}R}-\frac{{\rm i}}{k}\left(k^2+\frac{m^2}{R^2}\right)B'_R
+{\rm Pm\,Re}\frac{(\omega+m\Om)}{k}B'_R+\nonumber\\
&& \qq +\frac{m}{kR}X_4-\frac{m}{kR}b\,u'_R-u'_R=0,
\label{dbzdr}
\end{eqnarray}
\begin{eqnarray}
&&\frac{{\rm d}X_4}{{\rm d}R}-\left(k^2+\frac{m^2}{R^2}\right)B'_\phi
-{\rm iPm\,Re}(\omega+m\Om)B'_\phi +\nonumber\\
&& \qq + {\rm i}\frac{2m}{R^2}B'_R +2\frac{B}{R^2}u'_R-2{\rm Pm\,Re}\frac{b}{R^2}B'_R
+\nonumber\\
&& \qq + {\rm i}\frac{mB}{R^2} u'_\phi +{\rm i}k\,u'_\phi =0,
\label{dx4dr}
\end{eqnarray}
\beg
&&\frac{{\rm d}B'_R}{{\rm d}R}+\frac{B'_R}{R}+{\rm i}\frac{m}{R}B'_\phi+{\rm i}kB'_z=0
\label{6}
\ende
with $P'$ as the pressure and 
\begin{equation}
X_2=\frac{u'_\phi}{R}+\frac{{\rm d} u'_\phi}{{\rm d}R},  
 \ X_3=\frac{{\rm d} u'_z}{{\rm d}R},
  \ X_4=\frac{B'_\phi}{R}+\frac{{\rm d} B'_\phi}{{\rm d}R}. 
\label{61}
\end{equation}
The boundary conditions for the flow are 
$u'_R=u'_{\phi}=u'_z=0$ for
$R=R_{\rm in}$ and 
$R=R_{\rm out}$.
The boundary conditions for the magnetic field are also straightforward;
taking the inner and outer cylinders to be perfectly conducting, they are
simply
$B'_R=0$, $ X_4=0$ for $R=R_{\rm in}$ and $R=R_{\rm out}$. 
We will consider only the particular radius ratio $\hat\eta=0.5$.


\section{Axisymmetric modes}
\begin{figure}[htb]
\psfig{figure=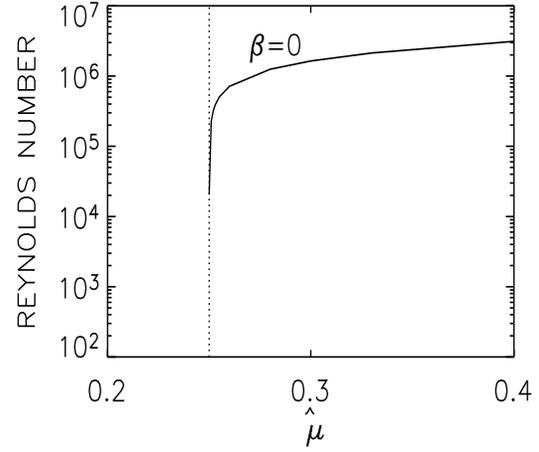,width=8.0cm,height=7.0cm}  
\caption{On the Rayleigh line (here $\hat \mu=0.25$, dotted line) the critical 
Reynolds number  scales as $\rm Pm^{-0.5}$, while beyond the Rayleigh line it 
scales as $\rm Pm^{-1}$, so that for ${\rm Pm}=10^{-5}$ the critical Reynolds number jumps from $10^4$ to $10^6$.}
\label{f1}
\end{figure}

\begin{figure}[htb]
\psfig{figure=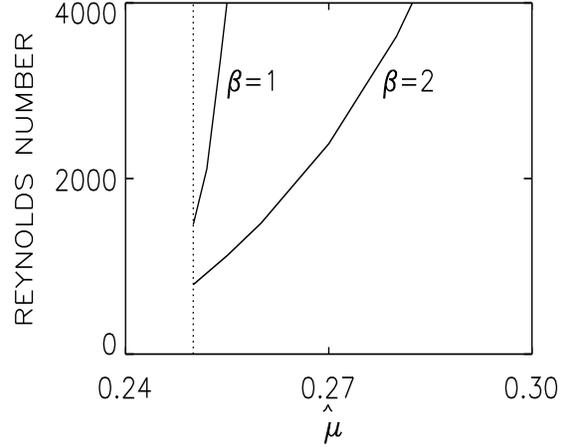,width=8.0cm,height=7.0cm}  
\caption{The critical Reynolds numbers beyond the Rayleigh line, for axisymmetric modes ($m=0$) with toroidal fields of the same order of magnitude as the axial
field. $\rm Pm=10^{-5}$.}
\label{f2}
\end{figure}

Figure~\ref{f1} shows the critical Reynolds numbers at and beyond 
the Rayleigh line for experiments without toroidal magnetic fields, the
classical design.  We note how $\rm Re_c$ jumps abruptly from $10^4$ to $10^6$
(for liquid sodium, see R\"udiger, Schultz \& Shalybkov 2003).  With a toroidal field, however, the solutions are very
different, as shown in Fig.~\ref{f2}.
The toroidal field strongly modifies the extremely steep line obtained for 
$\beta=0$ (Fig.~\ref{f1}). Its inclination is reduced and it starts at lower 
Reynolds numbers. For $\beta \gsim 2$ we find critical Reynolds numbers 
of order $10^3$. This is a dramatic reduction of the values of order $10^6$ which are characteristic for $\beta=0$.

\begin{figure}[htb]
\psfig{figure=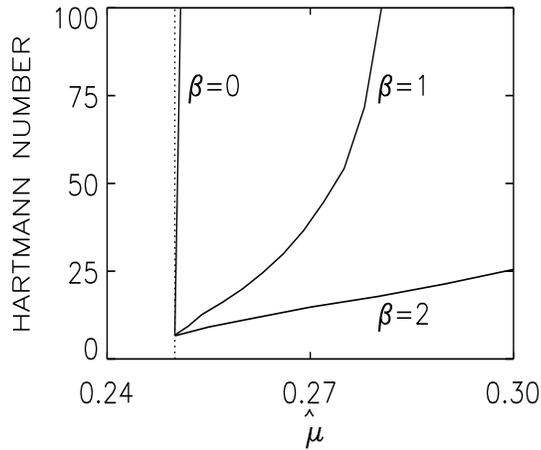,width=8.0cm,height=7.0cm}  
\caption{ The Hartmann numbers of the solutions given in Fig.~\ref{f2}. }
\label{f3}
\end{figure}

The corresponding Hartmann numbers for these instabi\-li\-ties are also
strongly reduced by the inclusion of a toroidal field (Fig.~\ref{f3}).
Tables~\ref{tab2} and \ref{tab2b} show results for $\hat\mu=0.27$,
$\beta=0$ to 4, and $\rm Pm=10^{-5}$ (sodium, Table~\ref{tab2}) and $\rm Pm=10^{-6}$
(gallium, Table~\ref{tab2b}).  The first five columns show the nondimensional
quantities indicated.  The last four columns show dimensional quantities,
taking $R_{\rm in}=5$ cm and $R_{\rm out}=10$ cm.   Specifically, Re has
been converted to $f_{\rm in}$, the rotation frequency, in Hz, of the inner
cylinder.  The Hartmann number Ha has been converted first to $B_0$, the strength, in Gauss, of the
axial field.  Next, the product $\beta\,$Ha has been converted to the strength
of the azimuthal field at $R_{\rm in}$.  Finally, $J$ denotes the axial
current, in Ampere, that is required to maintain this toroidal field.

\begin{table}
\tabcolsep5pt
\caption{\label{tab2} Characteristic values for $\hat \eta=0.5$, 
$\hat\mu=0.27$, and $\rm Pm=10^{-5}$.  In converting from nondimensional to
dimensional quantities $R_{\rm in}=5$ cm and $R_{\rm out}=10$ cm were used,
and the material properties of sodium. The magnetic fields are measured in Gauss, $B_\phi$ denotes the toroidal field at  the 
inner cylinder, the current $J$ is measured in Ampere. $\Re(\omega)$ is the real part of the Fourier frequency. }
\bigskip
\begin{tabular}{l|cccc|cccc}
\hline
  $\beta$  & Re & Ha & $k$ & $\Re(\omega)$ &
 $f_{\rm in}$ & $B_0$ & $B_\phi$ & $J$\\[0.5ex]
\hline\\[-8pt]
0 & $1\cdot 10^6$ &  542 & 1.7 & 0 & 45 & 883 & 0 & 0\\[0.5ex]
1&33833 &38.4 & 0.6& 0.04 & 1.5 &63&63&1565 \\[0.5ex]
2 & $2383$& 14.6 & 1.3 & 0.10 & 0.11  &24  &48  & 1190\\[0.5ex]
3 & 1160 & 10.7 & 1.6 & 0.13 &0.05 & 17 & 52 & 1308  \\[0.5ex]
4 & 842 & 9.5& 2.0  & 0.15& 0.04  & 15  & 62  & 1549 \\[0.5ex]
\hline
 \end{tabular}
 \end{table}
\begin{table}
\tabcolsep5pt
\caption{\label{tab2b}  The same as in Table \ref{tab2} but for gallium ($\rm Pm=10^{-6}$).}
\bigskip
\begin{tabular}{l|cccc|cccc}
\hline
  $\beta$  & Re & Ha & $k$ & $\Re(\omega)$ & $f_{\rm in}$ & $B_0$
 & $B_\phi$ & $J$\\[0.5ex]
\hline\\[-8pt]
0 & $1 \cdot 10^7$ &  1720 & 1.7 & 0&200  &7568 &0 & 0\\[0.5ex]
1&38250&39 & 0.6&0.04&0.8& 171 &171 &4290\\[0.5ex]
2 & 2382& 14.6 & 1.3 & 0.10& 0.05  &64  &128  & 3212\\[0.5ex]
3 & 1160 & 10.8 & 1.7 &0.13&0.02&48&143&3564\\[0.5ex]
4 & 842 & 9.5& 2.0  & 0.15& 0.02  & 42  & 167  & 4180 \\[0.5ex]
\hline
 \end{tabular}
 \end{table}

For $\beta\gsim2$ we see then that the critical Reynolds numbers no longer scale
as $\rm Pm^{-1}$, as they do for $\beta=0$, but rather become independent of Pm.
This is the same result previously noted by Hollerbach \& R\"udiger (2005) for
insulating boundaries.  More quantitatively, we note that for $\beta\gsim2$
the required rotation rates of the inner cylinder are reduced to less than
1 Hz, the axial fields to a few tens of Gauss, and the axial currents to
$O(1000)$ A.  All of these values are easily achievable in the laboratory.
We conclude therefore that this new design, incorporating an axial current,
is the most promising design for obtaining the MRI in a laboratory experiment.

\begin{figure}[htb]
\psfig{figure=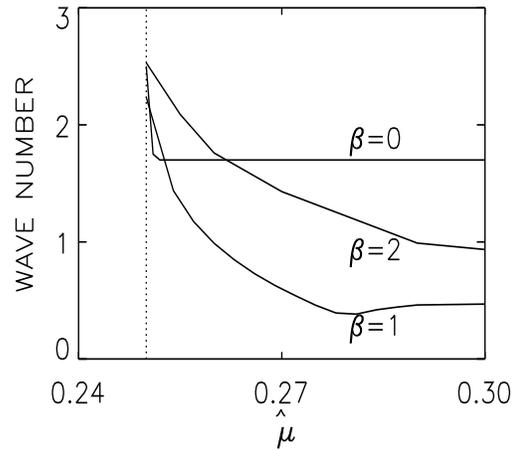,width=8.0cm,height=7.0cm}  
\caption{The wave numbers $k$ of the solutions given in Fig.~\ref{f2}.}
\label{f4}
\end{figure}

It is important also to consider the wave numbers $k$; if the vertical cell
size is too large the experiment will still run into difficulties.  We note
in Fig.~\ref{f4} that for non-zero $\beta$ $k$ is reduced somewhat,
corresponding to a greater extent in $z$ (the vertical cell size is given by
$\delta z\approx\pi/k$).  The increase in cell size is not very great though,
and therefore should not cause any problems.

Another interesting point in comparing $k$ for $\beta=0$ versus $\beta>0$ is
the Pm-dependence, or rather the lack thereof, since $k$ is independent of Pm
in both cases.  The details of how this comes about are subtly different though.
For the classical MRI with $\beta=0$ the critical wave number is
proportional to ${\Om_{\rm in}}/{V_{\rm A}}$, where
$V_{\rm A}=B_0/\sqrt{\mu_0 \rho}$ is the Alfv\'{e}n velocity.  The
nondimensional wave number $k$ then becomes
\begin{equation}
k\propto \frac{{\rm Re}\sqrt{\rm Pm}}{{\rm Ha}},
\label{kstern}
\end{equation}
which does {\emph{not}} depend on the magnetic Prandtl number, since $\rm Re
\propto Pm^{-1}$ and $\rm Ha\propto Pm^{-1/2}$.
In contrast, for $\beta>0$ $k$ is still independent of Pm, but so are Re and Ha,
so that (\ref{kstern}) cannot be the relevant balance in this case.  For $\beta>0$
the actual wave numbers are much greater than (\ref{kstern}) would predict.

\begin{figure}[htb]
\psfig{figure=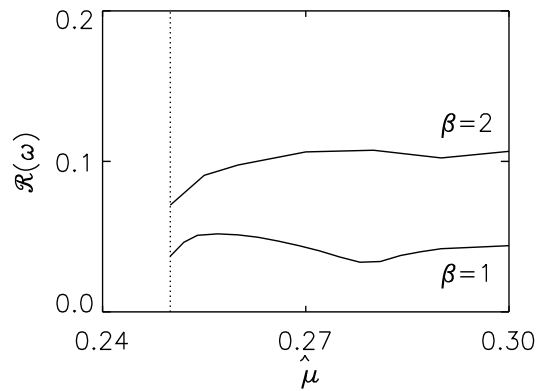,width=8.0cm,height=6.0cm}  
\caption{ The real parts of the eigenfrequency $\omega$
of the solutions given in Fig.~\ref{f2}.}
\label{f05}
\end{figure}

Finally, Fig.~\ref{f05} shows the frequency $\Re(\omega)$ of these modes.
For $\beta=0$ this is zero, since the modes are stationary in that case.
For non-zero $\beta$ stationary modes no longer exist; as noted also by
Hollerbach \& R\"udiger (2005), including a toroidal field changes the
symmetries of the problem in such a way that $\pm z$ are no longer equivalent,
which inevitably means that the modes will drift one way or the other in $z$,
that is, they will be oscillatory rather than stationary.  We see though that
the real parts of  $\omega$ are still rather small, with the mode-oscillation
time exceeding the rotation time of the inner cylinder by more than a factor
of 10. 

\begin{table}[h]
\caption{\label{tabold} Insulating cylinders: Fields and axial current for
$\hat\eta=0.27$ and $\beta=4$ ($R_{\rm in}=5$ cm, $R_{\rm out}=10$ cm).
The magnetic fields are measured in Gauss, $B_\phi$ denotes the toroidal
field at  the inner cylinder, the current $J$ is measured in Ampere.
}
\medskip
\begin{tabular}{l|cccccc}
\hline
&Re&$f_{\rm in}$ [Hz]&Ha & $B_z$  & $B_\phi$  & $J$ \\[0.5ex]
\hline\\[-8pt]
 sodium &1521& 0.07&16.3& 26 & 106 & 2657\\[0.5ex]
 gallium &1521&0.03 &16.3& 72 & 287 & 7170\\[0.5ex]
    \hline
\end{tabular}
\end{table}

\begin{figure}[h]
\vbox{
\psfig{figure=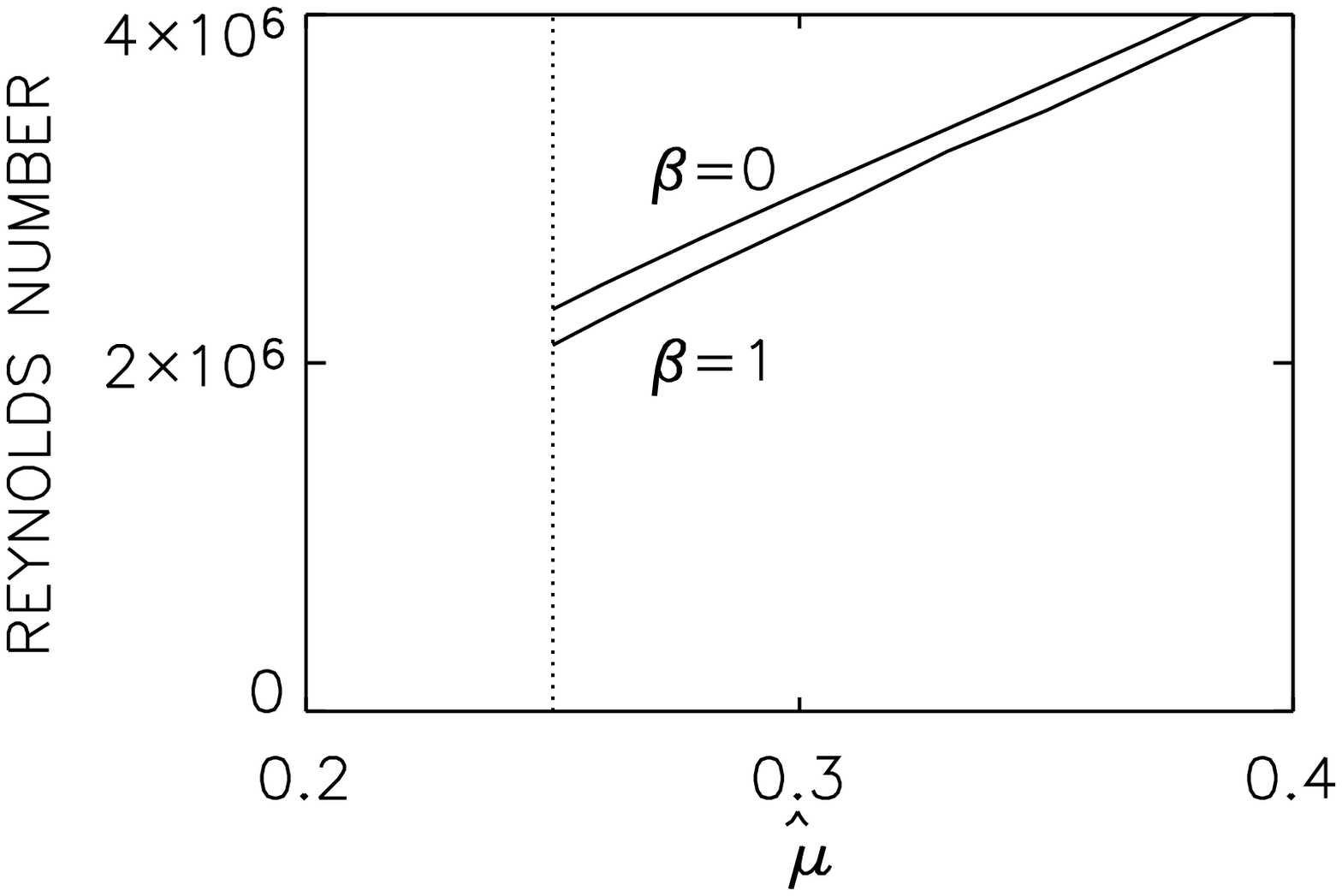,width=8.0cm,height=7.0cm}
\psfig{figure=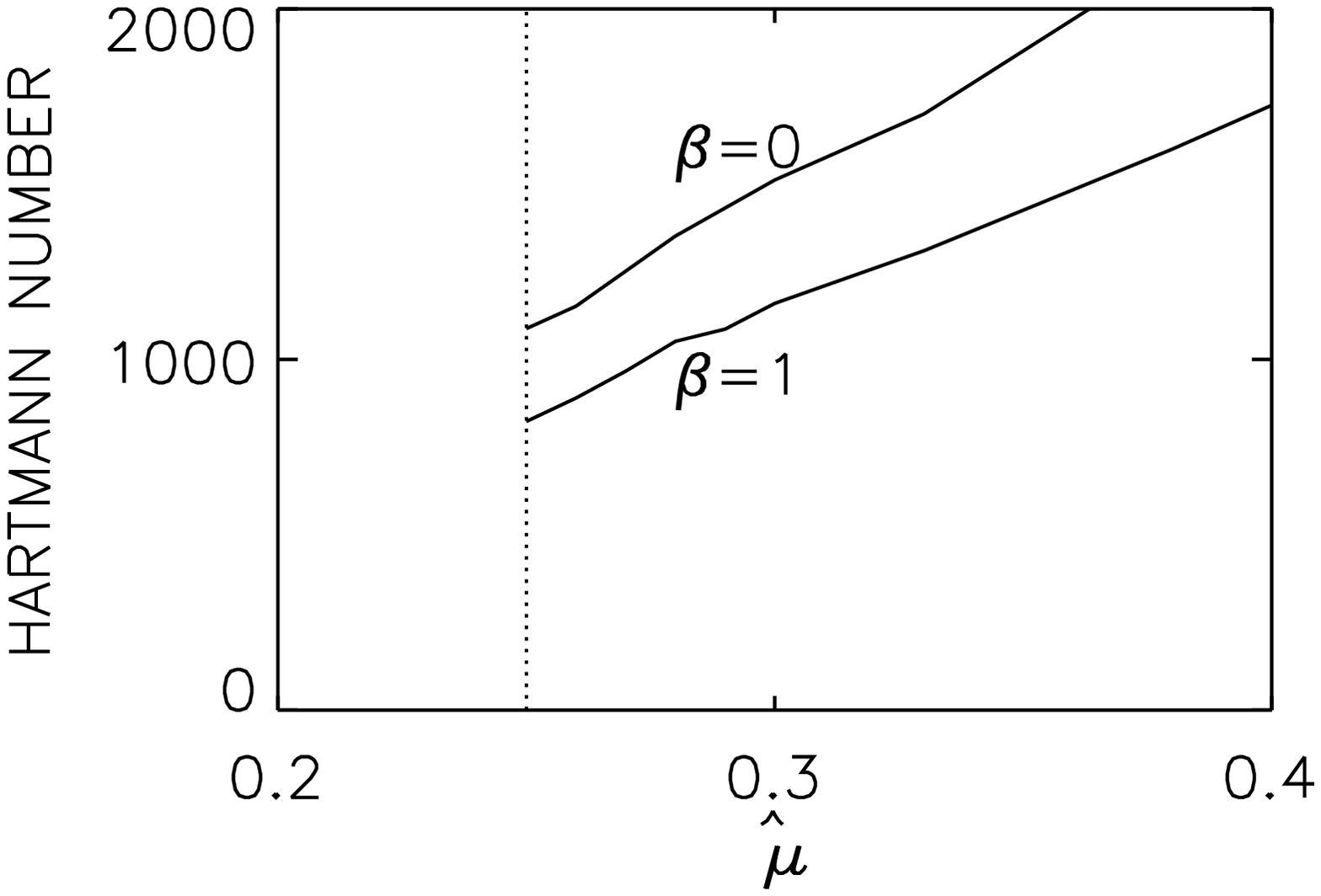,width=8.0cm,height=7.0cm}
\psfig{figure=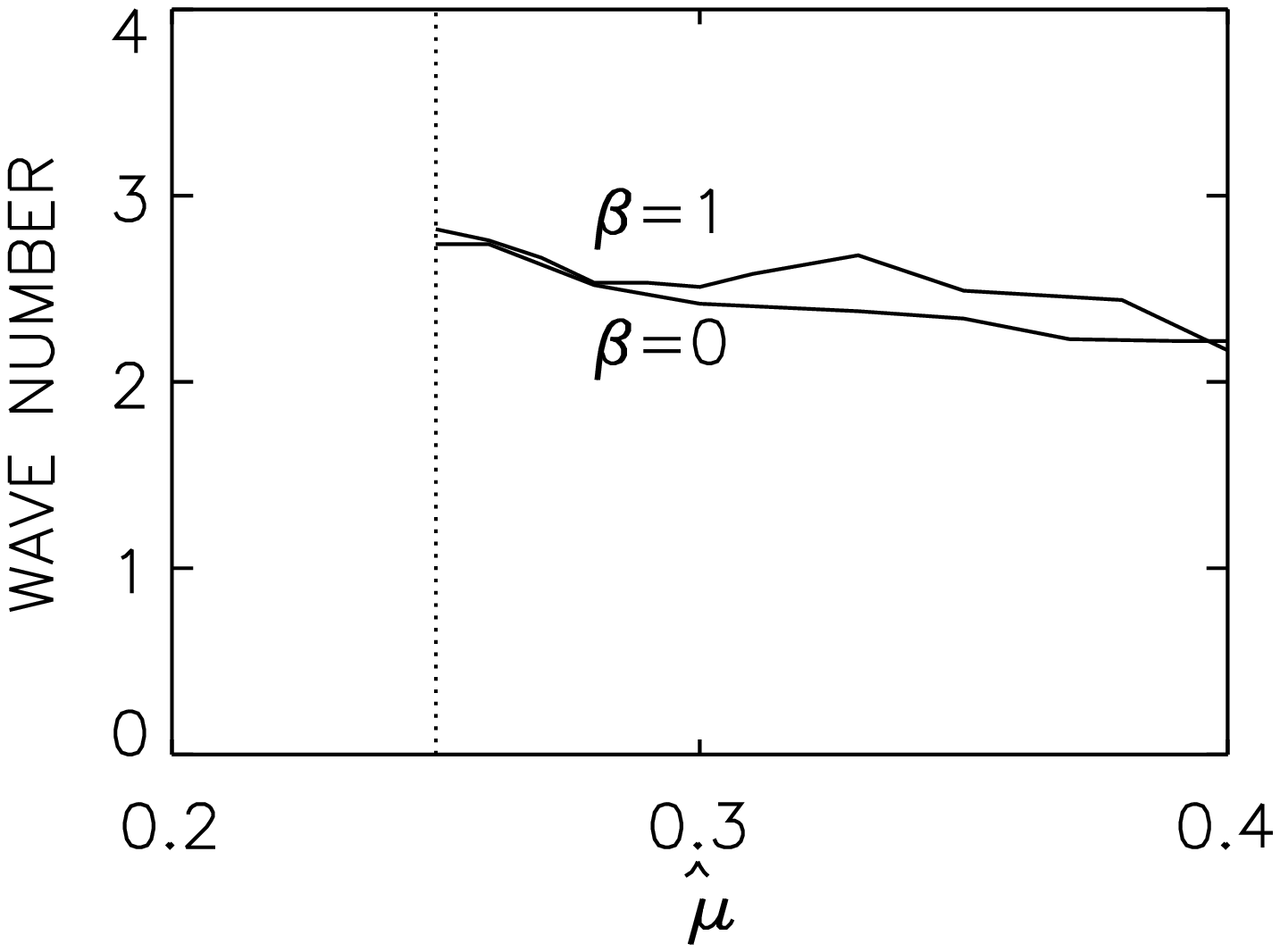,width=8.0cm,height=7.0cm}
}
\caption{Critical Reynolds  numbers (top),  Hartmann numbers (middle)  and wave
numbers (bottom) for nonaxisymmetric modes with $m=1$.}
\label{f5}
\rule{0pt}{60pt}
\end{figure}

\section{Nonaxisymmetric modes}
Hollerbach \& R\"udiger considered only axisymmetric modes $m=0$.  These
are usually the preferred modes, both in nonmagnetic Taylor-Couette flow, as
well as in the classical $\beta=0$ MRI.  Nevertheless, for a complete analysis
the nonaxisymmetric modes with $m>0$ must also be considered.  Figure~\ref{f5} shows
these results, and demonstrates that here too the critical Reynolds numbers for
the onset of nonaxisymmetric modes are greater than for the onset of axisymmetric
modes.  Indeed, including a toroidal field reduces $\rm Re_c$ by far less for
the nonaxisymmetric than for the axisymmetric modes.  For $\beta\gsim1$ the
axisymmetric modes thus occur several orders 
of magnitude before the nonaxisymmetric ones do.
These nonaxisymmetric modes are therefore not interesting from the
point of view of doing laboratory experiments.


\section{Conclusion}
The central conclusion of this work is as before in Hollerbach \&
R\"udiger (2005),
that imposing both axial and azimuthal magnetic fields together dramatically
reduces the critical Reynolds numbers required to obtain the magnetorotational
instability.  However, for the insulating boundary conditions considered there,
the axial currents required to generate this new toroidal field were at the
upper range of what could be achieved in the lab.  Here we therefore considered
conducting cylinders, and found that the required currents (and Reynolds numbers)
are  smaller.  For comparison, Table~\ref{tabold} presents the results
of Hollerbach \& R\"udiger for $\beta=4$; we see that $\rm Re=1521$ and $\rm Ha=16.3$,
compared with $\rm Re=842$ and $\rm Ha=9.5$ here (the last rows of Tables \ref{tab2} and
\ref{tab2b}).  Switching from insulating to conducting boundaries thus
reduces both Re and Ha by almost a factor of 2, which would certainly help in
achieving these toroidal fields in the lab.  We conclude therefore that
implementing this experiment with conducting boundaries is the most promising 
design for exploring the magnetorotational instability in the laboratory.

\afterpage{\clearpage}

\end{document}